\documentclass[11pt]{article}
\usepackage{amsfonts,amssymb,amsmath,amscd}
\usepackage[margin=1in]{geometry}
\usepackage{authblk}
\usepackage{xurl,hyperref}

\usepackage{bbm}
\usepackage{mathrsfs}
\usepackage{graphicx}

\usepackage{epstopdf}

\usepackage{wrapfig}
\usepackage[usenames]{color}
\definecolor{Red}{rgb}{0.9,0,0.0}
\definecolor{Blue}{rgb}{0,0.0,1.0}

\def\cC{\mathcal{C}}

\def\cI{\mathcal{I}}

\def\cQ{\mathcal{Q}}

\def\bE{\mathbb{E}}

\def\bP{\mathbb{P}}

\def\bR{\mathbb{R}}

\def\sS{\mathscr{S}}

\def\bfD{\mathbf{D}}

\def\bfW{\mathbf{W}}

\def\bfb{\mathbf{b}}

\def\bfw{\mathbf{w}}

\def\bfpi{\boldsymbol{\pi}}

\newcommand{\1}{\mathbbm{1}}            

\DeclareMathOperator*{\argmax}{arg\,max} 

\usepackage[inkscapelatex=false]{svg}

\title{
Groundwater Management: Combating the Sinking Feeling
}

\author{ 
	Igor Cialenco\,\thanks{Department of Applied Mathematics, Illinois Institute of Technology
		\newline \hspace*{1.45em}  10 W 32nd Str, Building RE, Room 220, Chicago, IL 60616, USA
		\newline \hspace*{1.45em}  Email: \url{cialenco@iit.edu},  URL: \url{http://cialenco.com}
        \newline \hspace*{1.45em} ORCID: \url{https://orcid.org/0000-0002-1825-0097}
		\vspace{0.5em}} 
	\and   
	 Michael Ludkovski\,\thanks{Department of Statistics and Applied Probability, University of California Santa Barbara
		\newline \hspace*{1.45em}  South Hall, Santa Barbara, CA 93106-3110, USA
		\newline \hspace*{1.45em} Email: \url{ludkovski@pstat.ucsb.edu}, URL: \url{http://ludkovski.pstat.ucsb.edu/}, 
        \newline \hspace*{1.45em} ORCID: \url{https://orcid.org/0000-0001-7887-3870}
		\vspace{0.5em}}
        }

\date{ {\small  First Circulated and This Version: August 21, 2026}}

\begin{document}

\maketitle

\begin{abstract}
 This non-technical survey introduces the topic of groundwater management to the mathematics and statistics communities. We focus on the emerging marketplaces for groundwater pumping rights, as well as outlining probabilistic and statistical approaches to groundwater dynamics that underpin stakeholder pumping decisions. After discussing rationing schemes and regulatory policies we list a range of open problems that are well suited to be tackled by the mathematical sciences researchers.  
\end{abstract}

\section{Shrinking Aquifers}

Groundwater---fresh water pumped from underground aquifers---is a critical source  in regions where precipitation and surface water resources are limited. For example, in California's Central Valley which supplies over 30\% of fruits and vegetables in US \cite{usgs-CentralValley}, groundwater accounts for 30-70\% of irrigation \cite{faunt2016water}, especially during droughts and rainless summers. Historically, groundwater was extracted through small wells and viewed as part of the landowners' prerogative. Large-scale farming and accompanying industrial-scale pumping has led to a severe case of the tragedy of the commons: without a regulatory mechanism, the ``free'' groundwater resource is over-extracted by individual users. The impacts of the unsustainable status quo are manifold: dry wells, land subsidence, saltwater contamination, permanent aquifer compression, surface stream collapse, etc. 

To bring order to the ``free-for-all'' runaway pumping, management mechanisms have been adopted, aiming  to track and regulate groundwater usage. To stabilize the aquifers, these  entail capping aggregate pumping, often substantially below business-as-usual levels.  Market-driven frameworks, essential to navigate the decades-old legal governance intricacies, are a major part of these solutions.  In this article we survey the nexus of emergent mathematical and statistical 
research themes linked to this topic. 

Groundwater management must contend with several challenges. First, aquifers form a complex hydrologic system that is only partially understood. Typically located several hundred feet under the surface, aquifer geometry is driven by rock composition and porosity. Hydrologists rely on indirect techniques to estimate and map out the volume of groundwater in a given basin. This inverse problem is compounded by the dynamic nature of groundwater that includes  downward percolation, lateral flows with varying hydroconductivity, and partially confined sub-basins. Hence the classical ``bathtub'' mental picture of an aquifer is inadequate and managers must account for numerous hydrological uncertainties and imprecise measurements. 

Second, groundwater use is governed by an extensive legal framework that is binding across the US and other common-law countries. Pumping is a privilege afforded either to  \emph{overlying} landowners (thus access is proportional to land claims) or to \emph{appropriative} stakeholders, i.e.~based on grandfathered claims. Unlike other free-access resources like air (or greenhouse gases and $CO_2$) or surface water, governments have limited direct ownership of aquifer resources and can only regulate bulk consumption. This makes groundwater both explicitly physical (for instance it is impossible for a third party to pump) and contingent on legislation.  Third, most agricultural basins naturally refill via deep percolation from surface precipitation. This is in contrast to exhaustible resources like hydrocarbons, or managed environments like fisheries. The variable groundwater recharge rate, driven by exogenous weather and climate patterns, creates seasonal-scale stochastic dynamics.

In view of the above, existing groundwater management practices \cite{jakemanEtAl2016Book} are rooted in an allocation paradigm, illustrated in the diagram below. First, pumping volumes are mandated to be measured \cite{maples2018leveraging} and strictly reconciled via water rights accounting. Second, regulators shape the overall pumping cap that modulates legal groundwater rights. Finally, regulators also set rules governing pumping rights exchange and carryover. Extensive literature supports the resulting market approach as the most effective for sustainable management \cite{ayres2021environmental,AyresEtAl2021,HanakEtAl2023}.
So far, key regions where groundwater markets have been established include California, Nebraska, Arizona, Texas, and South Australia. The most prominent example is California's 2014 Sustainable Groundwater Management Act (SGMA) that mandates the creation of Groundwater Sustainability Agencies (GSAs) in each basin. In turn, the GSAs are able to craft local, custom-designed markets for their stakeholders, as long as their Groundwater Sustainability Plans are regulator-approved.

\bigskip 
\includegraphics[width=0.8\linewidth, trim=0.2in 0.3in 0.1in 0in]{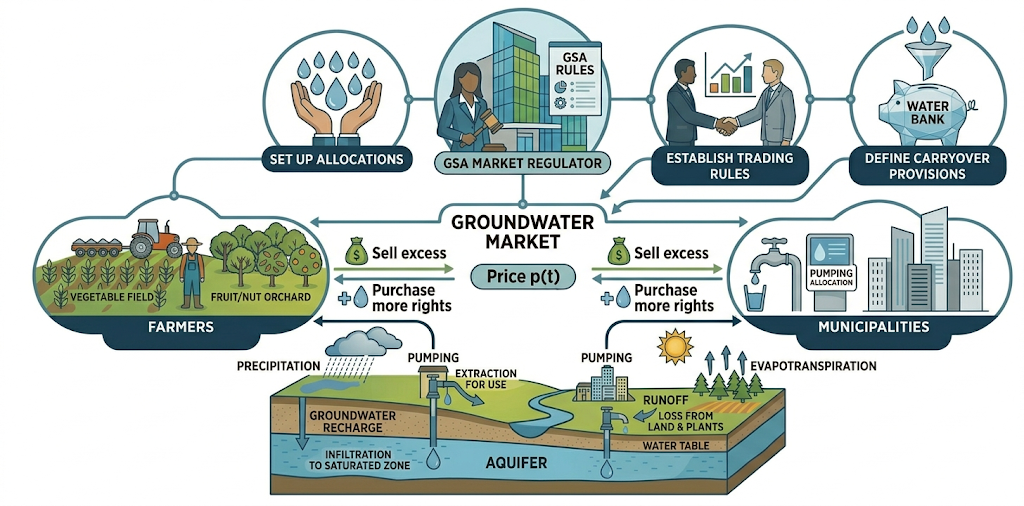}

\bigskip 

We emphasize the novelty of these marketplaces and the evolving policy making: fifty years ago, the idea that a farmer must meter his private well and possibly buy/sell the respective rights would be met with disbelief. 
From a modeling perspective, the outlined features highlight the twin pillars of multi-period optimization and stochastic game theory: (i) groundwater dynamics are stochastic, driven by precipitation and collective pumping decisions; (ii) the market is multi-period with the aquifer as nature's storage ``inventory''; (iii) stakeholders act competitively and interact through trading their groundwater rights, while being jointly impacted by the water table height and the exogenous weather and climate; (iv) the market is highly illiquid (pun intended) with  physical and trading constraints; (v) regulators focus on the overall sustainability of the basin, with further decisions delegated to individual users.   
    
\section{Statistical Approach to Groundwater Dynamics}

Groundwater resources are conventionally described through the height of the water table. Managers are rather concerned with  aquifer usable volume (ac-ft) $W$. The basic accounting is that $W$ is increased through recharge $R$ and is depleted through pumping $C$, and lateral outflow $O$. Viewing all these quantities dynamically, indexed by year $t$, we have $W(t+1) = W(t)+R(t+1)-C(t)-O(t)$. 

As mentioned, $C(t)$ is now closely monitored, and $O(t)$ is also trackable. In contrast, recharge is a latent quantity, best linked to the observable surface precipitation $P(t)$. Taking into account spatial heterogeneity, recharge at location $x$, $R(t;x)$ is to first-order a linear function of $P(t;x)$ with the proportionality constant referred to as specific yield $S_Y(x)$ and determined by soil composition and geologic layers. To estimate recharge over a sub-basin, Lithologic Texture Models combine records of thousands of driller logs to build a digital grid $ x \mapsto S_Y(x)$.  

An extensive and active literature studies physical groundwater dynamics, straddling hydrology (geologic properties of rock layers) and environmental science (mapping atmospheric conditions and climatic patterns into precipitation and surface flows, and then mapping these into aquifer recharge). The current gold standard are Integrated Water Budget approaches that rely on massive numerical models, based on partial differential equations and dynamical systems, to give a high-fidelity digital representation of aquifer evolution. Such models, e.g., the  California Central Valley Fine-Grid Groundwater Surface Water Simulation (C2VSimFG) tool \cite{C2V},  simulate the entire hydrologic cycle,  accounting for river seepage, mountain block recharge (water moving underground from the Sierras), and deep percolation.  The respective $S_Y(\cdot)$ values are calibrated by statistically matching model-based simulations against historical water table data, with analysis augmented with InSAR (Interferometric Synthetic Aperture Radar) and GRACE satellites that measure millimeter-scale land elevations, correlating land subsidence and rebound with the volume of water recharged.  

\begin{figure}[!htbp]
    \centering
    \includegraphics[width=0.8\linewidth,trim=0.05in 0.2in 0.05in 0in]{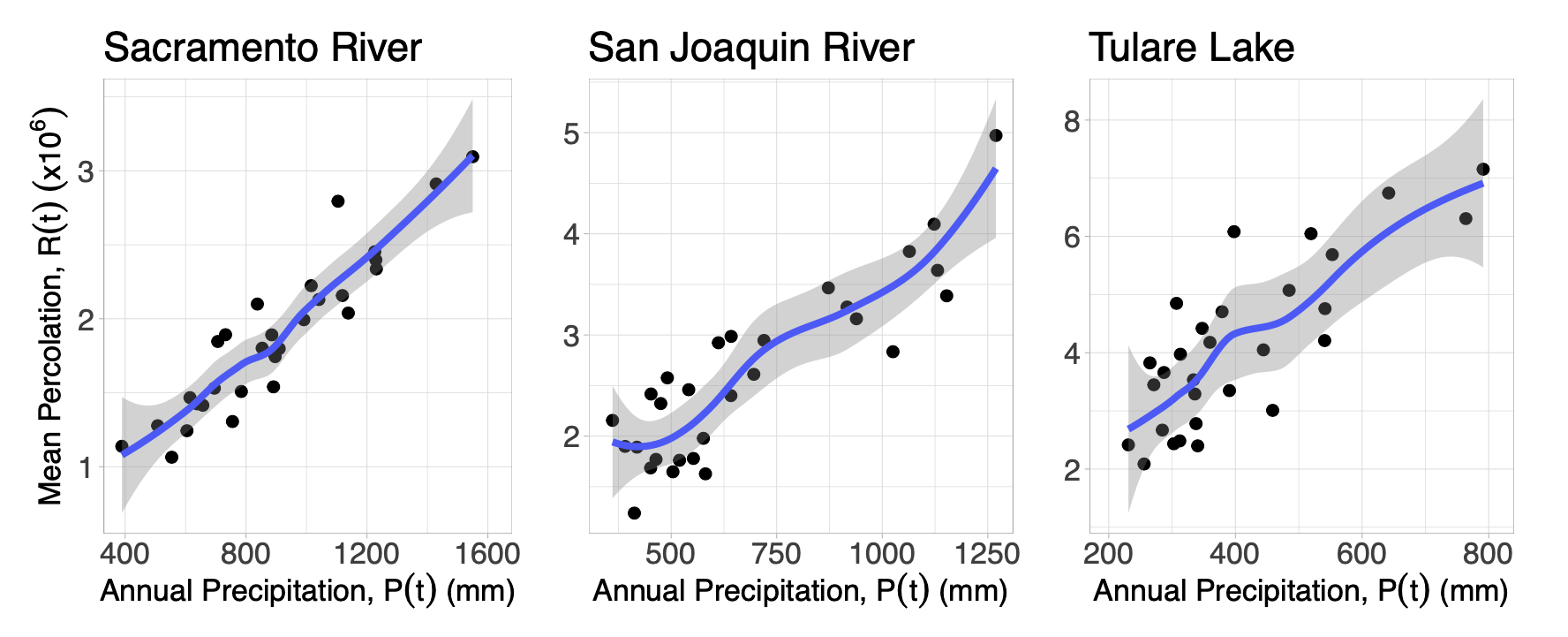}
    \caption{LOESS regression of mean percolation (y-axis) vs annual precipitation in 3 California basins.}
    \label{fig:prcp}
\end{figure}

As a check, Figure \ref{fig:prcp} shows a LOESS regression of $R(\cdot)$ against $P(\cdot)$ in several California basins. The linear pattern in the middle of the plots is supplemented by an S-shape: with very low precipitation, minimal deep percolation occurs, conversely with very heavy precipitation there is soil saturation and increased surface run-off. To run the regression, we sourced precipitation data from the California Department of Water Resources (DWR) Open Data Portal which offers a spatially gridded (through statistical post-processing) 100-year historic daily weather time-series. The estimated groundwater recharge data is provided by C2VSimFG which outputs model-based monthly deep percolation volumes across 21 hydrological sub-regions. LOESS estimates a smooth regression function $R(t) = \mu(P(t))+\varepsilon$ of the form $\mu(P)=\sum_{\ell=0}^2 \beta_\ell(P)P^\ell$ by learning the pointwise coefficients  $\beta_\ell(P)$ via minimizing the weighted least squares  objective $\sum_i W( \frac{p_i-p}{h})(r_i-\sum_\ell \beta_\ell (p_i-p)^\ell)^2$ over the training data $\{(p_i, r_i)\}$. The localized fit is adaptively imposed through the finite-support kernel (tricube function) $W$ and its bandwidth $h$.

Recently, these statistical reconciliation methodologies have burst into public view due to sanctions and counter-lawsuits between California State Water Board and Central Valley farmers\footnote{One can follow along via local news articles:
\begin{itemize}
    \item  9/13/2024: \url{https://sjvwater.org/kings-county-judge-rules-against-state-water-board-in-high-stakes-groundwater-case/}
\item 2/12/2026: \url{https://sjvwater.org/state-supreme-court-declines-to-hear-groundwater-case-out-of-kings-county/}
\item 6/5/2026: \url{https://sjvwater.org/kings-county-judge-considering-whether-to-advance-groundwater-lawsuit/}, 
\item the original DWR sanction decision justification that mentions InSAR on land subsidence:
\url{https://www.waterboards.ca.gov/sgma/docs/tule/202604-tule-exclusion-request-pres.pdf}
\end{itemize}}. The former alleges that the $S_Y$ values used in some GSPs are too optimistic, inappropriately inflating recharge amounts. Specifically, the State points to discrepancies between InSAR measurements and claimed percolation volumes, and inconsistencies regarding specific yields applied for adjacent parcels in different GSAs as a reason to put some areas under probation.  Because the math regarding $S_Y$ and land subsidence is so contentious, the state is now requiring more flow meters on individual wells, manifesting that ``you cannot manage what you cannot measure''.

For management and policy aspects, the above paradigm is over-engineered. To isolate the dynamic uncertainty, a probabilistic, non-hydrological setting is more suitable. Since decisions are made seasonally, the critical feature is to model the annual recharge time series. For instance in California, $R(t)$ is driven by the outcomes over the Water Year (Oct 1-April 30) as there is minimal precipitation during the Summer.
Abstracting from the geophysical characteristics, a natural reduced-form structure  for $\{P(t)\}$ is a Markov chain (a discrete-time memoryless process):
\begin{align}\label{eq:markov}
\bP( P(t+1) = i | P(t) = j) = Q_{ij}, \qquad i,j \in {\cal Q},
\end{align}
fully characterized by its transition matrix $Q$. The underlying precipitation ``regimes'' ${\cal Q}$ can be linked to  meteorological patterns, reflecting, e.g., the El Nino/La Nina cycles that drive seasonal West Coast rainfall, or estimated statistically through hidden Markov model techniques \cite{gobet2025interpretable}. The dynamics \eqref{eq:markov} embed the common mixture models for $P(t)$ and can be tractably extended to non-stationary climate. 
Coupled with the regression paradigm as in Figure \ref{fig:prcp}, which translates from $P(t)$ to $R(t)$, \eqref{eq:markov} yields a coherent, tractable reduced-form stochastic weather generator that emphasizes variable groundwater recharge (as opposed to deterministic projections that are still sometimes used but are woefully shortsighted). In the example below, we simplify further and identify $R(t)$--which is a continuous quantity---with $P(t)$, the discrete recharge regime.

\section{Competitive Market Model}

Armed with a probabilistic description of $R(t)$ and $W(t)$, we next describe a mathematical abstraction for the groundwater market. 
Consider $J$ economic agents (farmers, municipalities, landowners) who are stakeholders within a given basin or GSA. These agents trade groundwater rights among themselves, on a discrete time grid $t=0,1,\ldots, T$, using the same set of information.    At the start of each period, each  agent $j$ gets an allocation of groundwater rights $R_j(t)$. This could be a fixed or random amount, in particular dependent  on the groundwater recharge process $\{R(t)\}$ modeled as an exogenous Markov chain as in the previous section.

Agents make production decisions, such as which crops to grow and on how many acres, according to a profit-and-loss function $G_j$, which maps ac-ft of water use $C_j\geq 0$ by agent $j$ to her net profit (in dollars) $G_j(t,C_j)$ \cite{SearsEtAl2019}. We assume that $G_j$ is continuous and increasing in $C_j$---more pumping raises profits. 

At the beginning of each period $t$ trades occur, with $\psi_j(t)$ denoting the ac-ft of water traded, by agent $j$, with  the convention that  $\psi_j>0$ means selling water, and $\psi_j<0$ buying it. These trades are ``annual leases'' and only apply for the given period.  Denoting by $p(t)$, the price (in \$/ac-ft)  of groundwater at time  $t$, 
the $j$-th agent profit is given by 
\begin{align}\label{eq:L}
L_j(t) := G_j(t,C_j(t)) + \psi_j(t) \cdot p(t). 
\end{align}
Groundwater trades must balance out via the \textit{market clearing condition} 
\begin{equation}\label{eq:mrktClearPsi}
\sum_{j=1}^J \psi_j(t) = 0,  
\end{equation}
imposed as a binding and hard constraint. Thus, the amount of water $W_j(t)$ available for use by agent $j$ in period $t$ follows the dynamics 
\begin{equation}\label{eq:dynWj}
W_j(t+1) = W_j(t) + R_j(t+1) - C_j(t) - \psi_j(t), 
\end{equation}
for $t=0,\ldots, T-1$, with $W_j(0) = w_j$.

Rudimentary water accounting is based on seasonal pumping rights $R_j(t)$ and hence decomposes the multi-period horizon into a sequence of discrete harvest periods. However, stakeholders take a long view both in terms of necessary investments and because much groundwater-dependent agriculture involves perennial crops, such as orchards. Thus, mitigating seasonal variability, is top of mind for farmers and water districts. To facilitate such multi-year management,  many groundwater markets offer opportunities for \emph{banking} that allows agents to carry over their water rights to the next year, $C_j(t)+\psi_j(t)< W_j(t)$. Banking enables intertemporal smoothing of consumption: during wet years reserves are built up, while during droughts the carryover can be used up to maintain steady irrigation.
Denote the banked water by  agent~$j$ in period $t$ as
\begin{align}\label{eq:bj}
b_j(t) := W_j(t) - C_j(t) - \psi_j(t),
\end{align}
satisfying  $0\leq b_j(t)\leq \sum_{j}{W}_j(t)$. The non-negativity of $b_j(t)$ is essential: agents cannot borrow from their future allocations, but must first build up pumping reserves.

Given $W_j(t)$, agent $j$ controls dynamically her use of water $C_j(t)$, her trading $\psi_j(t)$, and her banked water $b_j(t)$. The pair $\pi_j(t) :=(C_j(t),\psi_j(t))$ is referred to as a policy of agent $j$ at time $t$, taking values in some compact set in $\bR^2$. 
We denote by $\sS_j$ the set of all feasible \emph{strategies} $\{\pi_j(t), t=0,\ldots, T-1\}$, understood as stochastic processes that are adapted to the flow of information. These are closed loop controls, meaning that the system dynamics are endogenously modified, as control actions depend on the current state and thus feed back into the evolution of the system. Moreover, the action $\pi_j$ of each agent  depends on the actions $\pi_{-j} := \{\pi_i, i\neq j\}$ of all other agents.  
Feasibility involves constraints that guarantee that $W_j(t) > 0$, for all $j$, and  $t$.

Each agent maximizes a risk-reward functional $U_j(\cdot)$ of her revenue.  The functions $U_j:\bR\to\bR$  are required to be monotone increasing and concave and account for agents' idiosyncratic attitudes to risk. Given the feasible  strategies  $\pi_{-j}$ chosen by others, and the price process $p$, agent $j$  
maximizes 
\[
A_j(\pi_j, \pi_{-j},p) = \bE^{\bfpi}\Big[\sum_{s=0}^T U_j(L_j(s; \pi_j(s),p(s))) \Big], 
\]
where $\bE^{\bfpi}$ denotes the expectation induced by the strategy profile  $\bfpi = (\pi_{1}, \ldots, \pi_{J})$. 
This expected profit depends both on the actions of all agents and on the evolution of the underlying  stochastic system $(\bfW(t), R(t))$, with the corresponding transition probabilities  constructed\footnote{For further details on this construction, we refer to the literature on Markov Decision Processes (MDPs); cf. \cite{Puterman1994}.} similarly to  \eqref{eq:markov}. 

To describe groundwater price formation as an outcome of an equilibrium, we  introduce a fictitious \emph{price-setter}, whose payoff at time $t$ is 
$L_{J+1}(\bfpi(t), p(t)) : = - p(t)  \big(\sum_{j=1}^J  \psi_j(t)\big)^2$.
The price-setter chooses $p(t)$, within a  finite range $0< \underline{p}<\overline{p} <\infty$. As such, $\{p(t)\}$ is a stochastic process, the dynamic maximizer of 
$$
A_{J+1}(p, \bfpi) := \bE^{\bfpi} \Big[\sum_{s=0}^T L_{J+1}(\bfpi(s),p(s))\Big], 
$$
which can be shown to yield a solution with $A_{J+1} \equiv 0$, so that \eqref{eq:mrktClearPsi} is satisfied.

While each agent acts in her self-interest, there are two levers that couple the overall market. First, the market-clearing condition couples the $\psi_j$'s via \eqref{eq:mrktClearPsi}. To enforce this constraint in equilibrium we have introduced the price-setter---interpreted as the regulator who has no criteria of their own beyond matching volumes of water that the agents wish to trade. Second, since agent allocations' $(w_1,\ldots, w_J)$ impact the resulting groundwater price, individual banking decisions end up affecting \emph{all} stakeholders next period, both directly through their expected profits and indirectly by modifying the set of future feasible actions.  The resulting  competition is framed as a \emph{non-cooperative} stochastic game, where agents optimize their pumping, trading and banking decisions while strategically aware of respective actions of others, and of uncertain future conditions. 
In line with the non-cooperative formulation, market outcomes are described through
a Nash Equilibrium (NE)\footnote{Actually a generalized NE since the feasibility set depends on aggregate pumping.}  \cite{BasarOlsder1999}
$(\bfpi^*(t), p^*(t))$, $t=0,\ldots,T$, meaning that $\forall j=1,\ldots, J,\forall \pi_j' \in \sS_j, p' \in \sS_{J+1}$, 
\begin{equation}\label{eq:Nash2}
\begin{split}
A_j(\pi_j^*, \bfpi_{-j}^*,p^*) &\geq A_j(\pi_j', \bfpi_{-j}^*,p^*), \\
A_{J+1}(p^*, \bfpi^*) & \geq A_{J+1}(p', \bfpi^*).
\end{split}
\end{equation}
Saying differently, a NE  is a strategy profile in which no agent can gain by unilaterally deviating. From the general theory of stochastic dynamic Nash games \cite{BasarZaccour2018}, one can show that there exists a NE in the class of mixed\footnote{By analogy to usual (pure) strategies, a mixed strategy for agent $j$ is a probability distribution on the action set $\cC_j(t)$.} strategies. 

Usually, this equilibrium is not unique. In \cite{CialencoLudkovski2025} it is shown that even for a one period setup, for every price $p>0$ there exists  a (pure) NE. It is therefore desirable to refine the problem and develop a method that yields a unique, or at least clearly computable, NE. One canonical method in selecting a NE is via Pareto optimality. 
A feasible strategy $(\bfpi,p)$ is called Pareto optimal if there is no other feasible strategy $(\bfpi',p')$ such that 
\[
\forall j \ A_j(\bfpi', p') \geq  A_j(\bfpi, p),   \exists i \ 
A_i(\bfpi', p') >  A_i(\bfpi, p). 
\]

\begin{figure}[!htbp]
    \centering
    \includegraphics[width=0.8\linewidth,trim=0.1in 0.2in 0in 0.1in]{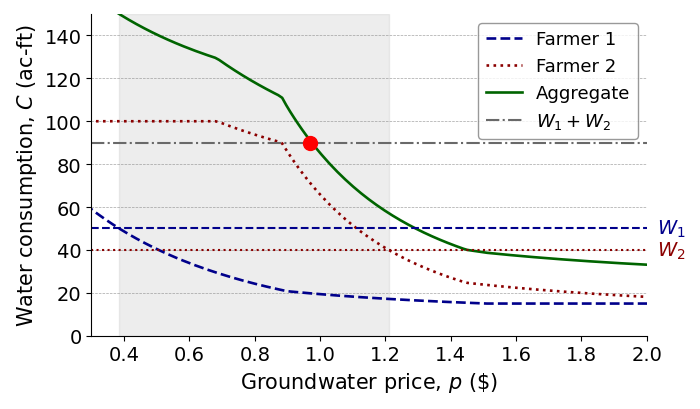}
    \caption{One period market model with two agents $J=2$. Shaded interval indicates the prices $[\underline{p}, \bar{p}]$ for which trades occur. Pareto price $p^\circ=0.975$ (red dot) corresponds to  $\sum_j C_j(p^\circ) = \sum_j W_j$. Horizontal dotted lines indicate the allocations $W_1, W_2$. }
    \label{fig:1period2payers}
\end{figure}

In \cite{CialencoLudkovski2025}, see Figure~\ref{fig:1period2payers}, we show that in a one period model, the Pareto strategy corresponds to the unique price $p^\circ$ at which the maximum amount of water is traded, and at which each agent achieves her maximal profit. Moreover, for any other price $p\neq p^\circ$, there are multiple trading NE scenarios, yielding different gains for individual agents.  Characterizing all NEs in a \emph{multi-period} setting---and selecting an economically meaningful one---is an open problem that is both theoretically important and computationally challenging. In practice, this is addressed on a case-by-case basis by exploiting the specific structure of the problem. Similar to classical stochastic control, one approach is to rely on the dynamic programming principle (DPP), which decomposes the multi-period game into a sequence of simpler games solved via backward induction. This corresponds to a sub-game perfect Markov NE \cite[Chapter 6.3]{BasarZaccour2018}, which can be equivalently described by a backward induction procedure, tentatively in a unique way, and usually only in the class of mixed Markov strategies.

For the proposed model, the DPP for sub-game perfect equilibria can be reduced 
to finding a fixed point of the best-response banking functions $b_j(t, \bfw, r; \bfb_{-j})$ that determine the banking of agent $j$ given available water profile $\bfw$ and others' banking strategies $\bfb_{-j}$. In the second step, the period-$t$ trading is resolved contingent on the equilibrium banking amounts $\bfb^*$, i.e.~given the 1-period model with water profile $\bfW(t)-\bfb^*(t,\bfW(t), R(t))$.

\begin{figure}[!htbp]
\centering
\includegraphics[width=0.8\linewidth,trim=0.1in 0.3in 0in 0in]{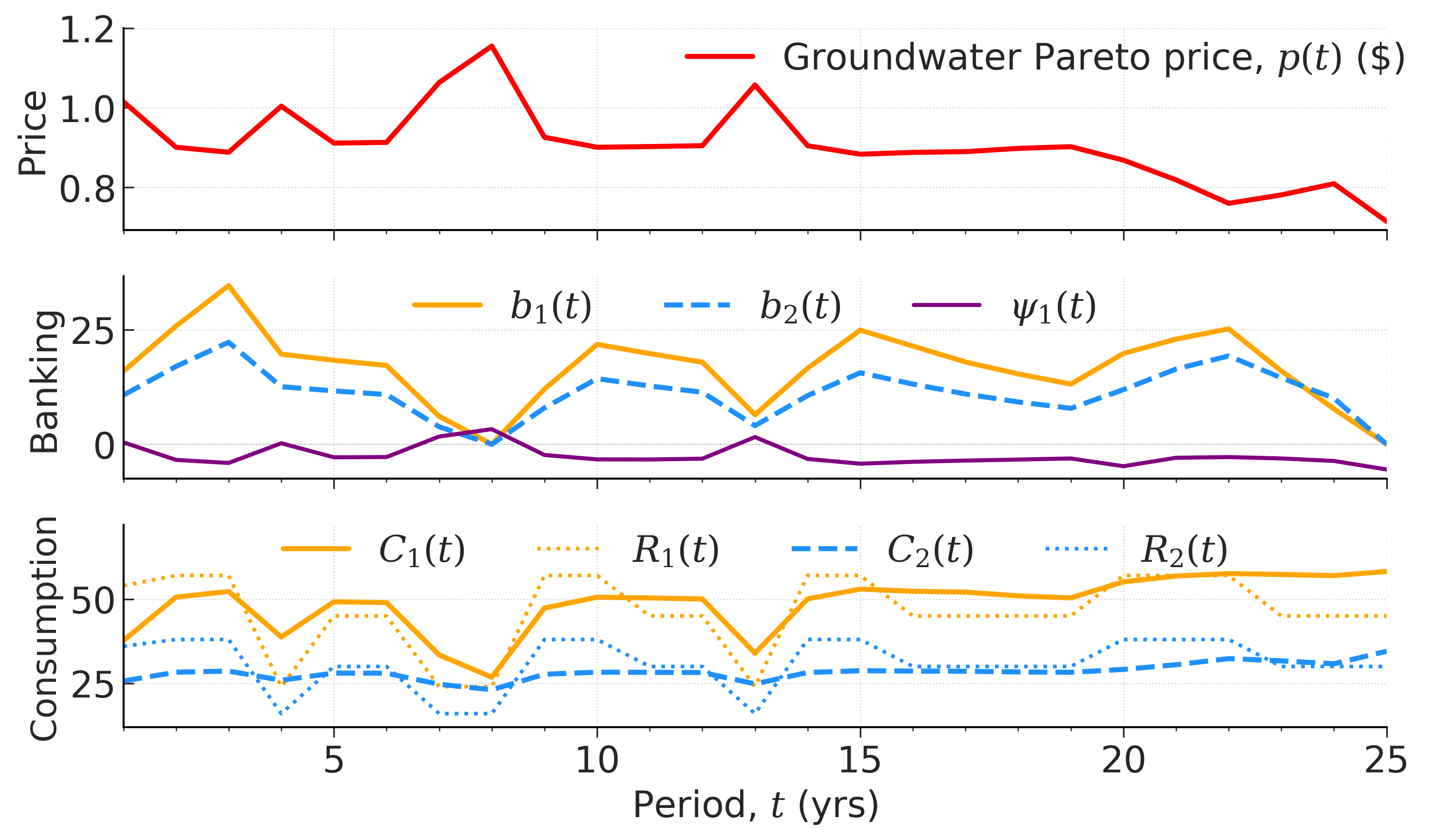}
\caption{Groundwater market scenario over 25 periods, showing a sample path of the prices $p(t)$ (top),  banking amounts $b_j(t)$ and trading amounts $\psi_1(t)$ (middle), as well as consumptions $C_j(t)$ and allocations $R_j(t)$ (bottom), in a $J=2$-sector market.  \label{fig:20-period}}
\end{figure}

Figure \ref{fig:20-period} based on the work in \cite{CialencoLudkovski2026} shows how banking supports intertemporal smoothing in a groundwater market with stochasticity modeled by a 3-state Markov chain for the recharge $\{R(t)\}$.
Thanks to banking, the consumption paths $C_j(\cdot)$ are nearly constant across the years (especially for Farmer 2) and much less variable than the per-period allocations $R_j(t)$ which are proportional to $R(t)$. When the recharge is low,  prices rise and the farmers draw on their banked reserves.  Consumption is further stabilized through trading: notably Farmer 1 buys water from Farmer 2 during normal years $\psi_1(t) < 0$, but sells it during droughts. Despite the simple dynamics of $\{R(t)\}$, market evolution is path-dependent and exhibits multiple dynamic phenomena. An astute reader would also note that carryovers and prices collapse in the last few periods $t>20$ due to the zero imposed terminal condition that nullifies any rights unused by $T$. This is a side-effect of our model choices and implies that we primarily want to study the quasi-stationary behavior at intermediate periods.  Other insights drawn from numerical experiments (not shown here) include the role of allocation rules, the auto-correlation of recharge (modifying $\cQ$), and the risk preferences $U_j(\cdot)$. For instance, more risk-averse agents will bank more in order to have a thicker reserves cushion against droughts. Similarly, more volatile climate induces a stronger (on average) demand for pumping reserves. 

In the outlined probabilistic model, the key uncertainty is from the future water rights $R_j(s), s>t$ which are coupled through joint dependence on the common shock $R(s)$. Additional consideration is  the complementarity between surface and ground flows, whereby a hot summer has the combined effect of decreasing stream flows, increasing irrigation needs due to evapotranspiration and lowering groundwater recharge, making $G_j(\cdot)$ dependent on $R(t)$. Moreover, stakeholders are also influenced by long-term effects such as changing climate, generational sustainability goals and multi-year actions (e.g., the wait to harvest a newly planted orchard). 

\section{Rationing Schemes}

Viewed as a scarce resource, groundwater allocation corresponds to a rationing scheme. In the above fully market-driven set up, the groundwater price is used as a clearing mechanism, monetizing the marginal value of pumping. A Pareto allocation then implies that the price $p^\circ$ is chosen to equate the volume of water bought and sold. Under this mechanism, less-efficient (in terms of their individual shadow price of water) agents will sell their rights to the more-efficient agents that extract higher utility from water. 

As soon as we have a market where $p \neq p^\circ$, the question of rationing re-appears. For example, suppose that regulators set a high groundwater price; then there would be more sellers than buyers and a scheme is needed to allocate the few rights that the ``most motivated'' buyers wish to purchase across all these sellers. One standard rule is based on pro-rata  \cite{CLT26-prorata} and has been used extensively in environmental rationing (timber and fishing quotas, pollution permits) and in financial markets (electronic trading of bonds). 

For simplicity, consider a one-period market with a regulator-assigned price $p$ and allocations $W_j$. Farmers optimize their profits, obtaining their desired water consumptions 
\[
C_j^\circ(p) = \argmax_{\underline{c}_j \leq C_j \leq \bar{c}_j} [G_j(C_j) + (W_j-C_j)p].
\]
The resulting total water trading demand $\bfD_{-}$ and supply $\bfD_{+}$ are denoted by
\begin{align*}
\bfD_\pm(p) := \pm\sum_j (W_j-C_j^\circ(p))\1_{W_j-C_j^\circ(p) \lessgtr 0}.  
\end{align*}

\begin{figure}[!htbp]
    \centering
    \includegraphics[width=0.8\linewidth,trim=0in 0in 0in 0.23in,clip=True]{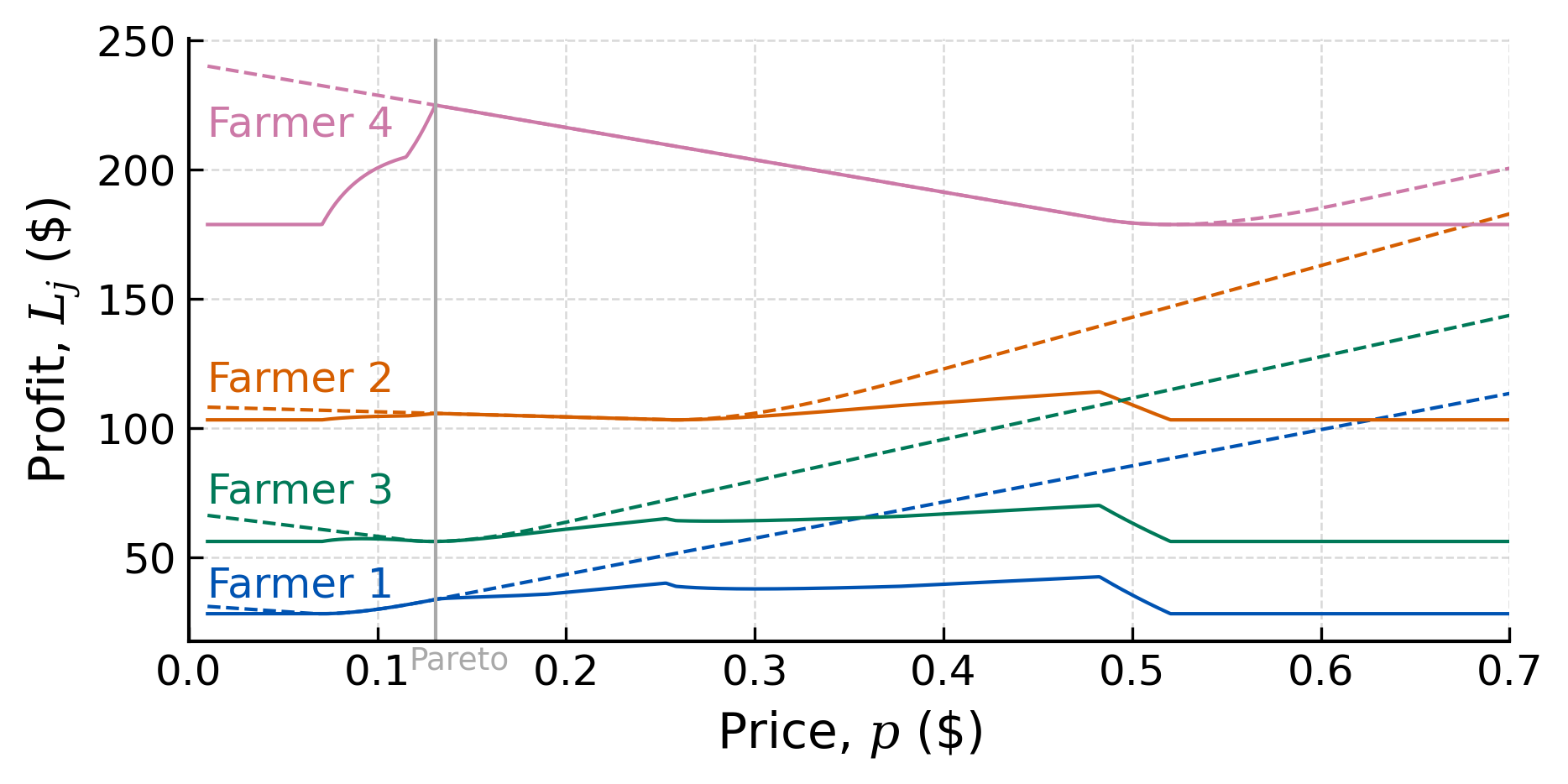}
    \caption{Profits under pro-rata allocation $L_j(\tilde{C}_j(p))$ (solid lines) and individually optimal consumption $L_j(C^\circ_j(p))$ (dashed lines) as a function of price $p$ for a groundwater market with $J=4$ farmers. }
    \label{fig:pro-rata1}
\end{figure}

Under a pro-rata algorithm the water traded by an agent is proportional to their individual demand/supply relative to the total demand/supply. For example, when supply $\bfD_{+} > \bfD_{-}$ is larger than demand,  each seller $j$ gets to sell $(W_j-C_j^\circ)\bfD_-/\bfD_+$, with unsold water used for production of goods,  up to the maximum amount $\bar{c}_j$. Thus, each seller $j$ will consume 
\[
\widetilde{C}_{j}(p) = \!\Bigl(C_{j}^\circ(p) + \frac{ W_j - C_j^\circ(p)}{\bfD_+(p) } 
( \bfD_+(p) - \bfD_-(p)) \Bigr)\wedge \bar{c}_j. 
\]
With such pro-rata allocations, cf.~Figure \ref{fig:pro-rata1}, the profits of a given agent are no longer necessarily monotone in $p$, and hence local changes to groundwater prices might have unintuitive impacts. 

A different pro-rata allocation scheme is to settle demand–supply imbalances in proportion to the assigned rights $W_j$. For example,  if $\bfD_+ > \bfD_-$, then seller $j$ will sell the amount    
\[
\cI_j(p):=\frac{W_j}{\sum_\ell W_\ell\1_{W_\ell \geq C^\circ_\ell(p)}} \cdot \bfD_+(p)\wedge (W_j-C_j^\circ(p)),
\]
and consume up to $(C_j^\circ(p) + W_j-\cI_j(p))\wedge \overline{c}_j$. 
Further alternatives include seniority-based rationing (i.e., based on a pre-determined ordering of the agents, for example in increasing order of their allocations $W_j$ to prioritize trades by small farms) and rationing trading (e.g., a cap on sales of excess water that is proportional to $W_j$, $\psi_j \le a_j W_j$)  \cite{CLT26-prorata}. Some rationing schemes require a round-based allocation mechanism: for example uniform rationing first divides total demand equally among all sellers $b_1:=\bfD_+/(\sum_j \1_{W_j-C_j^\circ(p) \geq 0}) $,  allowing each seller to trade up to $b_1$, i.e. $(W_j-C_j^\circ(p))\wedge b_1$. Sellers with unmet demand then proceed to the second round where the same process is repeated, and so forth. While pro-rata is transparent and avoids over-concentration,  it also creates moral hazard to inflate reported demand. An open question is how to define equitable rationing that accounts for multiple objectives.

\section{Regulatory Policy}
A primary aim of regulating groundwater is to mitigate the tragedy of the commons by accounting for various collective externalities   and balancing the interests of all stakeholders, not just those with the legal rights \cite{bruno2024designing,kuwayama2013regulation}.  As two examples, regulators seek to prevent irreversible aquifer compression due to land subsidence, and to ensure environmental protection of local communities  that are  legally subordinated to the pumping rights of farmers \cite{ayres2021environmental}. Below we sketch several extensions and enhancements of the proposed model which serves as a flexible baseline. Mathematical exploration regarding the properties of the resulting NEs and computational aspects should be viewed as open problems to  be treated separately for each sub-case.

In the Pareto setup, the  state authority  acts like an ``invisible hand'', effectively optimizing the aggregate profits of the market participants, without criteria of their own. Taking into account the above additional objectives, one approach is to view the regulator as optimizing $\sup_p [{\cal R}(p)-{\cal C}(p)]$, where ${\cal R}(p)$ is the ``revenue'' associated with groundwater price $p$, for example a weighted sum of farmers' gains and the value of environmental protection, and ${\cal C}(p)$ is the ``cost''. For instance, if we interpret ${\cal C}(p) \ge 0$ as the forgone trading gains relative  to the Pareto first-best market, we can connect to the pro-rata market above. 

\textbf{Policy Tools.} 
A market  where the price process $\{p(t)\}$ is first selected by the market-maker, and then farmers optimize their actions is an example of a leader-follower game. In the broader class of \emph{principal-agent} games there is a similar hierarchy where the regulator comes up with a market design, while strategically anticipating how stakeholders would react to such rules. In the second stage, pumpers interact among themselves, e.g., by trading their water rights.
In the previous section, the regulatory policy is parametrized by the water price  (set through taxation, subsidies, etc) that steers the pumping to a particular NE.  Below we outline further non-price-based mechanisms:
\begin{itemize}
\setlength{\itemsep}{0pt}

    \item Accounting for environmental or community priorities by modifying allocation rules or carryover provisions. This effectively creates a virtual stakeholder that is endowed with some rights and is part of the market. 
    
    \item Contingent rule making:  regulators invoke the possibility of ``taking over'' and slashing quotas in case desired collective outcomes are not achieved. Such punitive threats act as ``boundary conditions'' to coerce agents into adjusting their behavior to avoid the penalties. The state of California presently seeks to impose mandatory pumping cutbacks on GSAs that are SGMA-non-compliant.

    \item Trading by the regulator $\psi_{J+1}(\cdot)$, for instance selling reserves when there is drought at higher price than farmers trade among themselves. This is equivalent to the regulator acting as a backstop by allowing 
    farmers to pay a penalty for using extra water.

    \item Collective regulation based on the aggregate water table height. For instance, regulators might enact a \emph{ratchet} whereby allocations decrease as soon as the overall water table height drops below a threshold. Mathematically, this adds an additional coupling among the agents and forces them to internalize the cost of others' pumping;

    \item Localized trading or pumping constraints that aim to avoid land subsidence through excessive/imbalanced pumping in some sub-regions. This introduces network effects, converting the setup into a game on a graph;

    \item A cap-and-trade-like system of pumping credits, such as the Long Term Storage Credits in Arizona; new credits can be created by purposefully injecting surface (river) water underground and then ``extinguished'' during shortage years.  
\end{itemize}

Last but not least, it should be evident that there are many stochastic drivers that could be considered, such as stochastic water needs, uncertain harvests or stochastic prices. Including these leads to other forms of coupling between individual and collective uncertainties.

\textbf{Heterogeneous Stakeholders. }
Heretofore, we concentrated on markets with a small number of homogeneous, fully competitive stakeholders, matching the structure of most pumping districts. Two significant modifications concern (i) heterogeneity; (ii) coalition-formation. For the latter, agents may have incentives to band together, either to mitigate losses due to competition (the ``price of anarchy'') or to exercise market power. The rise of mega-farms (often owned by outside investors) that coordinate pumping across tens of thousands of acres is partly driven by this motive. Market behavior can shift perceptibly when there is a ``major'' player who has, for example, the capital to buy the rights of all the other agents. Making the market coalition- or collusion-proof is an important consideration in market design; so is mitigating concentration of water rights, as measured, say, by a Gini index.  In complement, some pumpers can commit to act cooperatively, e.g., to explicitly value environmental protection. 

As another instance of heterogeneity,
many basins straddle urban-rural settings and pit municipal (Muni) water districts against agricultural (Ag) stakeholders. These constituencies consume and trade water differently:
farmers aim to maximize their profits and may sell water rights by pumping less. Municipalities on the other hand must meet (at minimum cost) users' demand each period; they have no immediate demand elasticity, but may adjust their tariffs to impact consumption in the medium term. Moreover, Muni stakeholders are generally much larger than individual farms and face a distinct set of externalities 
Regulators tend to impose different prices for Ag-Ag and Ag-Muni transactions, or to limit respective Ag-Urban trades  to support the long-term agricultural sector viability.

\subsection{Multiple Data Basins}

Empirical analysis of groundwater management requires the collation of datasets from multiple disciplines, and with a few exceptions there is a dearth of  datasets that are amenable to statistical and corpus analysis \cite{maples2018leveraging}. Best organized are precipitation datasets,  curated and maintained by government resource agencies, such as the NWS. In contrast, gridded hydrologic data is not universally accessible, and typically consists either of some sparse (both spatially and temporally) estimates or model-based calibrations, such as the aforementioned C2VSim tool \cite{C2V, zeff2021california}. 

Data on cultivated and irrigated acreage, yields, agricultural prices and profitability must be downloaded from government Departments of Agriculture and agricultural economics datasets \cite{SearsEtAl2019}. A treasure trove of California-related data can be found in the Groundwater Sustainability Plans (GSPs) that each GSA must submit to the state. Often numbering 1000+ pages, the GSPs offer an integrated econo-hydro-political summary of a particular basin. The reports put out by the Public Policy Institute (PPIC) is another great resource \cite{AyresEtAl2021,HanakEtAl2023}. 

Actual pumping consumptions and histories of groundwater rights transactions are kept by the Water Masters. California alone numbers 100+ such record keepers, who have a wide latitude regarding the scope of official records and public disclosure; e.g., most of sale \emph{prices} are non-public. The recent availability of LLM- and Agentic AI tools opens the door for a large-scale effort to systematically digitize (as many records still consist of scanned pages or figures) and bring into a common database those sources. Indeed, improvements in the analysis of transaction data, even for a single water district, would represent a substantial achievement.  

\section{Problems to Unearth}

Groundwater management offers a unique setting where mathematical scientists are well positioned to be meaningful contributors, acting as the go-betweens for hydrologists, economists, policy analysts and climate scientists.  This pressing societal challenge is inherently regional and sidesteps the partisan baggage and geopolitics associated with carbon emissions, fossil fuels or even surface water. Moreover, the problem is structurally stochastic and demands deep understanding of the underlying uncertainties and risk trade-offs. Coupled with well-defined cost-benefit considerations, groundwater management is well-suited for optimization-based solutions. At the same time, there are numerous mathematical challenges to be tackled---from algorithms for finding equilibria in heterogeneous markets to addressing informational asymmetry among farmers, to geo-spatial estimation of groundwater flows.

The ongoing repercussions of the SGMA rules in California present a real-life testbed. 
The outlined multi-period framework provides a support tool to simulate market behavior under a range of market designs, yielding quantitative insights into potential outcomes. In parallel, dozens of GSAs are running real-time experiments in how to build robust groundwater regulations.

In \cite{CialencoLudkovski2026} the paradigm focused on a small number of agents and ongoing work concerns the extension to more realistic markets with many agents. Two promising research directions are Reinforcement Learning (RL) and Mean Field Games (MFG) \cite{perolat2017multi,hu2023recent}. 
The MFG framework  can be interpreted as a fixed land area being subdivided among ever more users, leading to a continuum of homogeneous infinitesimally small farmers. In this mean-field regime, decisions are understood as measures. For instance, the decision to carryover rights translates into a ``banking density'', summarized as the fraction of agents that are banking and the collective banked volume. A related major/minor $(J,A_M)$-setup features a large farmer who has access to $A_M$ acres and $J$ minor ones who each own $A_m$, with the total acreage $A_M + J A_m$ again fixed. The ultimate vision is an agent- and RL-based ``market simulator'' that can test out a variety of  regulatory configurations, such as using auctions to allocate rights. 

Another wide-open line of research concerns
network and multivariate phenomena that are important to regulators, such as  (i) penalties on water traded depending on distance between farmers; (ii) accounting for lateral water flows within/between basins; (iii) inter-basin or surface-groundwater trades. We also mention the ongoing discussions about \emph{managed recharge} whereby farmers are directly incentivized to flood their fields and fallowed land in order to purposefully maximize deep percolation. Some schemes offer additional pumping credits (net recharge rules) for doing so, partially endogenizing $\{R(t)\}$ dynamics.

 To be useful, all models of course must be calibrated, offering a range of challenges for statisticians to ponder, from how to capture the distribution of potential climate change impacts \cite{bruno2024designing}, to how to incorporate hydrological model error \cite{sears2022moment}. Closing the loop, there are causal statistics to be done estimating the impact of market design on water prices and transaction volumes.

\subsubsection*{Acknowledgments} 
IC acknowledges support from the National Science Foundation grant DMS-2407549. ML acknowledges support from the National Science Foundation grant DMS-2407550.

\bibliographystyle{siam}
\bibliography{WaterAllocation-2026-06-22,bib-new}

@TechReport{AyresEtAl2021,
  author      = {Andrew Ayres and Ellen Hanak and Brian Gray and Gokce Sencan and Ellen Bruno and Alvar Escriva-Bou and Greg Gartrell},
  institution = {Public Policy Institute of California (PPIC)},
  title       = {Improving {C}alifornia’s Water Market: How Water Trading and Banking Can Support Groundwater Management},
  year        = {2021},
  month       = {sep},
}

@TechReport{HanakEtAl2023,
  author      = {Hanak, Ellen and Ayres, Andrew and Peterson, Caitlin and Escriva-Bou, Alvar and Cole, Spencer and Joaquín, Zaira},
  institution = {Public Policy Institute of California (PPIC)},
  title       = {Managing Water and Farmland Transitions in the {S}an {J}oaquin Valley},
  year        = {2023},
  month       = {sep},
}

@Book{jakemanEtAl2016Book,
  author    = {Jakeman, Anthony J. and Barreteau, Olivier and Hunt, Randall J. and Rinaudo, Jean-Daniel and Ross, Andrew},
  publisher = {Springer Nature},
  title     = {Integrated groundwater management},
  year      = {2016},
  isbn      = {978-3-319-79502-7},
  pages     = {775},
  subtitle  = {Concepts, Approaches and Challenges},
}

@Article{maples2018leveraging,
  author    = {Stephen Maples and Ellen Bruno and Alejo Kraus-Polk and Stacy Roberts and Lauren Foster},
  journal   = {Water},
  title     = {Leveraging hydrologic accounting and water markets for improved water management: the case for a central clearinghouse},
  year      = {2018},
  month     = {nov},
  number    = {12},
  pages     = {1720},
  volume    = {10},
  doi       = {10.3390/w10121720},
  publisher = {MDPI},
}

@Article{ayres2021environmental,
  author    = {Andrew B. Ayres and Kyle C. Meng and Andrew J. Plantinga},
  journal   = {Journal of Political Economy},
  title     = {Do environmental markets improve on open access? Evidence from {C}alifornia groundwater rights},
  year      = {2021},
  month     = {oct},
  number    = {10},
  pages     = {2817--2860},
  volume    = {129},
  doi       = {10.1086/715075},
  publisher = {The University of Chicago Press Chicago, IL},
}

@Article{sears2022moment,
  author = {Sears, Louis S. and Lin Lawell, C.-Y. Cynthia and Torres, Gerald and Walter, M. Todd},
  title  = {Moment-based {M}arkov equilibrium estimation of high-dimension dynamic games: An application to groundwater management in {C}alifornia},
  year   = {2022},
}

@Article{kuwayama2013regulation,
  author    = {Yusuke Kuwayama and Nicholas Brozovi{\'{c}}},
  journal   = {Journal of Environmental Economics and Management},
  title     = {The regulation of a spatially heterogeneous externality: Tradable groundwater permits to protect streams},
  year      = {2013},
  month     = {sep},
  number    = {2},
  pages     = {364--382},
  volume    = {66},
  doi       = {https://doi.org/10.1016/j.jeem.2013.02.004},
  publisher = {Elsevier},
}

@Article{zeff2021california,
  author    = {Harrison B. Zeff and Andrew L. Hamilton and Keyvan Malek and Jonathan D. Herman and Jonathan S. Cohen and Josue Medellin-Azuara and Patrick M. Reed and Gregory W. Characklis},
  journal   = {Environmental Modelling \& Software},
  title     = {California's food-energy-water system: An open source simulation model of adaptive surface and groundwater management in the Central Valley},
  year      = {2021},
  month     = {jul},
  pages     = {105052},
  volume    = {141},
  doi       = {10.1016/j.envsoft.2021.105052},
  publisher = {Elsevier},
}

@manual{C2V,
author = "{California Department of Water Resources}",
year = 2026,
title = "Development and Calibration of the {C}alifornia {C}entral {V}alley Groundwater-Surface Water Simulation Model ({C2VSim}), Version 3.02-CG; Technical Memorandum",
note = "C2VSim Version 2025 available at https://data.cnra.ca.gov/dataset/c2vsimcg-version-2025-draft",
}

@article{hu2023recent,
  title={Recent developments in machine learning methods for stochastic control and games},
  author={Hu, Ruimeng and Lauri{\`e}re, Mathieu},
  journal={Numerical Algebra, Control and Optimization},
  volume={14},
  number={3},
  pages={435--525},
  year={2024},
   mrnumber  = "4795589",
  mrclass = {49-02 (49N80 60H30 68T07 90C40 91A15 93-08 93E20)},
}

@Book{BasarOlsder1999,
  author    = {Ba\c{s}ar, Tamer and Olsder, Geert Jan},
  publisher = {SIAM, Philadelphia, PA},
  title     = {Dynamic noncooperative game theory},
  year      = {1999},
  isbn      = {0-89871-429-X},
  series    = {Classics in Applied Mathematics},
  volume    = {23},
  mrclass   = {90-02 (90D10 90D25)},
  mrnumber  = {1657965},
  pages     = {xvi+519},
}

@article{gobet2025interpretable,
  title={Interpretable seasonal multisite hidden {M}arkov model for stochastic rain generation in {F}rance},
  author={Gobet, Emmanuel and M{\'e}tivier, David and Parey, Sylvie},
  journal={Advances in Statistical Climatology, Meteorology and Oceanography},
  volume={11},
  number={2},
  pages={159--201},
  year={2025},
  publisher={Copernicus Publications G{\"o}ttingen, Germany}
}

@article{SearsEtAl2019,
  author      = {Sears, Louis and Lim, David and Lin Lawell, C.-Y. Cynthia},
  title       = {Spatial Groundwater Management: A Dynamic Game Framework and Application to {C}alifornia},
  year        = {2019},
  month       = {Jan},
  number      = {01},
  doi         = {10.1142/S2382624X18500194},
  issn        = {2382-6258},
  journal     = {Water Economics and Policy},
  pages       = {1850019},
  publisher   = {World Scientific},
  volume      = {5},
}

@Book{BasarZaccour2018,
  editor    = {Ba\c{s}ar, Tamer and Zaccour, Georges},
  publisher = {Springer, Cham},
  title     = {Handbook of dynamic game theory},
  year      = {2018},
  isbn      = {978-3-319-44374-4; 978-3-319-44373-7},
  doi       = {10.1007/978-3-319-44374-4},
  mrclass   = {91-00},
  mrnumber  = {4139285},
  pages     = {xix+1285},
  url       = {https://doi.org/10.1007/978-3-319-44374-4},
}

@article{perolat2017multi,
  title={A multi-agent reinforcement learning model of common-pool resource appropriation},
  author={Perolat, Julien and Leibo, Joel Z and Zambaldi, Vinicius and Beattie, Charles and Tuyls, Karl and Graepel, Thore},
  journal={Advances in neural information processing systems},
  volume={30},
  year={2017}
}

@Article{CialencoLudkovski2025,
  author   = {Cialenco, Igor and Ludkovski, Mike},
  journal  = {SIAM J. Financial Math.},
  title    = {Short communication: a groundwater market model},
  year     = {2025},
  issn     = {1945-497X},
  number   = {2},
  pages    = {SC51--SC63},
  volume   = {16},
  doi      = {10.1137/25M1727758},
  fjournal = {SIAM Journal on Financial Mathematics},
  mrclass  = {91B76 (91A10 91B70 91B72)},
  mrnumber = {4925172},
  url      = {https://doi.org/10.1137/25M1727758},
}

@article{bruno2024designing,
  title={Designing water markets for climate change adaptation},
  author={Bruno, Ellen M and Jessoe, Katrina},
  journal={Nature Climate Change},
  volume={14},
  number={4},
  pages={331--339},
  year={2024},
  publisher={Nature Publishing Group UK London}
}

@Book{Puterman1994,
  author    = {Puterman, Martin L.},
  publisher = {John Wiley \& Sons},
  title     = {Markov Decision Processes: Discrete Stochastic Dynamic Programming},
  year      = {1994},
  isbn      = {9780470316887},
  month     = apr,
  doi       = {10.1002/9780470316887},
  issn      = {1940-6347},
    SERIES = {Wiley Series in Probability and Mathematical Statistics:
              Applied Probability and Statistics},
   MRCLASS = {90C40 (90-02)},
  MRNUMBER = {1270015},
}

@Article{CialencoLudkovski2026,
  author  = {Cialenco, Igor and Ludkovski, Mike},
  journal = {Preprint arXiv:2605.26363},
  title   = {Dynamic Groundwater Markets},
  year    = {2026},
}

@article{CLT26-prorata,
  author  = {Cialenco, Igor and Ludkovski, Mike and Tekam Fongouo, Gael Dimitri},
  title   = {Pro-rata mechanisms in groundwater markets},
journal = {Preprint arXiv:2608.00917},
  year    = {2026}, 
}

@misc{usgs-CentralValley,
  author = {{US Geological Survey: California Water Science Center}},
  title = {Central {V}alley Regional Overview},
  year = {Accessed August 4, 2026},
  howpublished = "\url{https://www.usgs.gov/centers/california-water-science-center/central-valley-regional-overview}"
  }

@article{faunt2016water,
  title={Water availability and land subsidence in the {Central Valley, California, USA}},
  author={Faunt, Claudia C and Sneed, Michelle and Traum, Jon and Brandt, Justin T},
  journal={Hydrogeology Journal},
  volume={24},
  number={3},
  pages={675--684},
  year={2016},
  publisher={Springer}
}

\end{document}